\documentclass[aps,prl,reprint,groupedaddress]{revtex4-1}

\newcommand{\be}{\begin{equation}}
\newcommand{\ee}{\end{equation}}

\usepackage{verbatim}
\usepackage{latexsym,amssymb,float}
\usepackage{setspace}
\usepackage{graphicx}
\usepackage{epsfig}
\usepackage{amsmath}
\usepackage{bm}

\begin{document}

\title{Differential Conductance and Defect States in the Heavy Fermion Superconductor CeCoIn$_5$}

\author{John S. Van Dyke$^1$}
\author{James C. Davis$^{2,3,4,5}$}
\author{Dirk K. Morr$^1$}

\affiliation{$^1$ University of Illinois at Chicago, Chicago, IL 60607, USA}

\affiliation{$^2$ CMPMS Department, Brookhaven National Laboratory, Upton, NY 11973, USA}
\affiliation{$^3$ LASSP, Department of Physics, Cornell University, Ithaca, NY 14853, USA}
\affiliation{$^4$ School of Physics and Astronomy, University of St. Andrews, Fife KY16 9SS, Scotland}
\affiliation{$^5$ Kavli Institute at Cornell for Nanoscale Science, Cornell University, Ithaca, NY 14853, USA}

\date{\today}

\begin{abstract}
We demonstrate that the electronic bandstructure extracted from quasi-particle interference spectroscopy [Nat. Phys. {\bf 9}, 468 (2013)] and the theoretically computed form of the superconducting gaps [Proc. Nat. Acad. Sci. {\bf 111}, 11663 (2014)] can be used to understand the $dI/dV$ lineshape measured in the normal and superconducting state of CeCoIn$_5$ [Nat. Phys. {\bf 9}, 474 92103)]. In particular, the $dI/dV$ lineshape, and the spatial structure of defect-induced impurity states, reflects the existence of multiple superconducting gaps  of $d_{x^2-y^2}$-symmetry. These results strongly support a recently proposed microscopic origin of the unconventional superconducting state.
\end{abstract}


\maketitle

CeCoIn$_5$ \cite{Pet01} has long been considered the ``hydrogen atom" of heavy fermion superconductivity \cite{Ste79,Miy86,Beal86,Sca86,Sca87,Lav87,Col89}, and much experimental \cite{Iza01,Mos01,Koh01,Cur12,Sto08,Park08,Hu12,Koi13,Tru13,Shu14,Kim15} and theoretical effort \cite{Fli10,Sca12,Das13,Dav13,Liu13,Yang14,Wu15,Ert15} has focused on illuminating its unconventional properties \cite{Kim01,Pag03,Bia03,Pag06,How11}, and the microscopic mechanism underlying the emergence of superconductivity. While a variety of experimental probes have reported evidence for the existence of nodes \cite{Iza01,Mos01,Koh01,Cur12}, a sign change of the superconducting (SC) order parameter along the Fermi surface \cite{Sto08,Park08}, and spin-singlet pairing \cite{Koh01,Cur12}, theoretical efforts \cite{Fli10,Sca12,Das13,Dav13,Liu13,Yang14,Wu15} to provide a quantitative or even qualitative explanation for the properties of the superconducting state in CeCoIn$_5$ have been hampered by insufficient insight into the material's complex electronic bandstructure \cite{Koi13}. This situation, however, has recently changed due to a series of groundbreaking scanning tunneling spectroscopy (STS) experiments \cite{Ern10,All13,Zhou13} which have yielded unprecedented insight into the electronic structure of the superconducting state, providing strong evidence for its unconventional $d_{x^2-y^2}$-wave symmetry. In particular, the analysis of quasi-particle interference (QPI) spectroscopy experiments \cite{All13} has provided a quantitative understanding of the momentum structure of the hybridized heavy and light bands, as well as that of the unconventional $d_{x^2-y^2}$-wave superconducting gap. This, in turn, has allowed the formulation of a microscopic theory ascribing the superconducting pairing mechanism to the strong magnetic interactions in the heavy $f$-electron band \cite{Dyke14}. Moreover, detailed measurements of the $dI/dV$ lineshape in the normal and superconducting states of CeCoIn$_5$ \cite{Zhou13} have revealed a series of puzzling observations. In particular, in the normal state, $dI/dV$ exhibits a feature which was hypothesized to be like that seen in the pseudo-gap state of the cuprate superconductors \cite{Lee06}. Furthermore, in the superconducting state, the $dI/dV$ lineshape possesses a kink at energies well inside the SC gap, and an intriguing spatial form in the vicinity of defects \cite{Zhou13}. These observations have raised three important questions: (1) What is the origin of the pseudo-gap like feature in $dI/dV$ in the normal state? (2) Does the kink in the $dI/dV$ lineshape and its spatial form near defects reflect the existence of multiple superconducting gaps with $d_{x^2-y^2}$-wave symmetry? (3) Can the electronic structure extracted from QPI spectroscopy \cite{All13} and the theoretically computed superconducting gaps \cite{Dyke14} be used to explain the
unconventional features observed in the $dI/dV$ lineshape \cite{Zhou13}?

In this article, we address all three questions. In particular, we demonstrate that the experimentally measured $dI/dV$ lineshape in the normal and superconducting states can be described using the electronic structure extracted from QPI spectroscopy \cite{All13} and the computed form of the superconducting gaps \cite{Dyke14}. This allows us to show (a) that the experimentally observed pseudo-gap like feature in the normal state of CeCoIn$_5$ arises from the momentum structure of the hybridized light and heavy bands in the heavy fermion state, (b) that the kink-like feature in the $dI/dV$ lineshape below $T_c$ reflects the presence of multiple superconducting gaps, and (c) that the experimentally measured spatial form of $dI/dV$ near defects is a signature of the unconventional $d_{x^2-y^2}$-wave symmetry of the superconducting gap. The good agreement between our theoretical results and STS experiments \cite{Zhou13} provides further validity for the electronic structure of CeCoIn$_5$ extracted in Ref.~\cite{All13} and the microscopic pairing mechanism proposed in Ref.~\cite{Dyke14}.

To investigate the form of the differential conductance, $dI/dV$,  in CeCoIn$_5$ in the normal and superconducting states, we start from the electronic bandstructure extracted from QPI spectroscopy \cite{Dyke14}, described by the mean-field Hamiltonian $H_{MF} = H_0 + H_{SC}$ where
\begin{subequations}
\begin{align}
H_{0} &=\sum_{{\bf k},\sigma} \left( \varepsilon^c_{\bf k} c^\dagger_{{\bf k},\sigma} c_{{\bf k},\sigma} +  \varepsilon^f_{\bf k} f^\dagger_{{\bf k},\sigma} f_{{\bf k},\sigma} + s_{\bf k} f^\dagger_{{\bf k},\sigma} c_{{\bf k},\sigma} + h.c \right) \nonumber \\
& = \sum_{{\bf k},\sigma} \left( E^\alpha_{\bf k} \alpha^\dagger_{{\bf k},\sigma} \alpha_{{\bf k},\sigma} + E^\beta_{\bf k} \beta^\dagger_{{\bf k},\sigma} \beta_{{\bf k},\sigma} \right) \label{eq:1a} \ ; \\
H_{SC} &= -\sum_{\bf k} \left( \Delta^\alpha_{\bf k} \alpha_{{\bf k}, \downarrow}\alpha_{{\bf -k}, \uparrow} + \Delta^\beta_{\bf k} \beta_{{\bf k}, \downarrow}\beta_{{\bf -k}, \uparrow} \right) \ . \label{eq:1b}
\end{align}
\end{subequations}
Here, $c^\dagger_{{\bf k},\sigma}, f^\dagger_{{\bf k},\sigma}$ create a particle with momentum ${\bf k}$ and spin $\sigma$ in the (light) conduction and heavy band, respectively, $\varepsilon^{c,f}_{\bf k}$ are the conduction and heavy band dispersions, and $s_{\bf k}$ is the effective hybridization between these two bands. The second line of Eq.(\ref{eq:1a}) results from a unitary transformation to new operators $\alpha_{{\bf k},\sigma},\beta_{{\bf k},\sigma}$ that diagonalizes $H_0$, resulting in the three Fermi surfaces shown in Fig.\ref{Fig:SCgap}(a). Moreover, $H_{SC}$ in Eq.(\ref{eq:1b}) is the mean-field BCS Hamiltonian reflecting intraband pairing within the $\alpha$- and $\beta$-bands (the detailed form of the band structure and superconducting gaps are given in Ref.~\cite{Dyke14}). Interband pairing is neglected due to the momentum mismatch between the three Fermi surfaces \cite{Dyke14}. The spatially resolved differential conductance at site ${\bf r}$ is then obtained via \cite{Fig10}
\begin{multline}
\frac{dI({\bf r},V)}{dV}= \frac{2\pi e}{\hbar} N_t t_c^2 [ N_c({\bf r},eV) + (t_f / t_c) ^2 N_f({\bf r},eV) \\ + 2 (t_f / t_c) N_{cf}({\bf r},eV) ]
\end{multline}
where $N_t, N_c, N_f$ are the densities of states of the STS tip, conduction electrons, and localized $f$-electrons, and $N_{cf}$ describes the correlations between the light and heavy bands \cite{Fig10}.  Here, the ratio $t_f / t_c$ controls the relative tunneling strength into the $f$- and $c$-electron bands.
\begin{figure}
 \includegraphics[width=8cm]{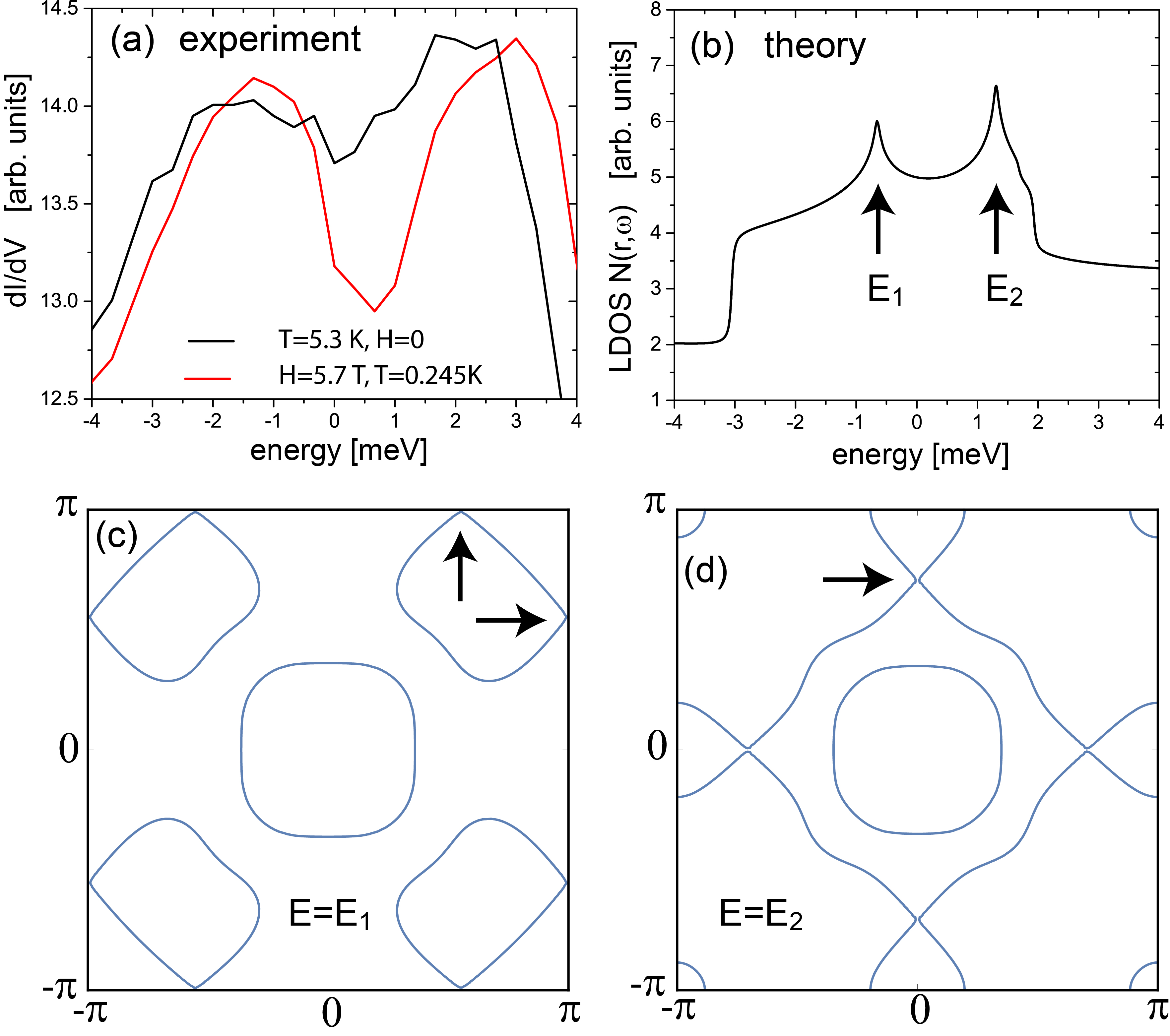}
 \caption{(a) Experimental $dI/dV$ curve in the normal state of CeCoIn$_5$ \cite{Zhou13}, courtesy of A. Yazdani. (b) Theoretically computed $dI/dV$ in the normal state with $t_f/t_c=0.1$ (c),(d) Equal-energy contours at $E_1$ and $E_2$, respectively. \label{Fig:dIdVnormal}}
 \end{figure}

We begin by discussing the form of the differential conductance in the normal state. Zhou {\it et al.} \cite{Zhou13} reported that above $T_c$, or when the SC state is suppressed by a magnetic field, $dI/dV$ exhibits a feature that might be related to a pseudogap, consisting of two peaks in $dI/dV$, as reproduced in Fig.~\ref{Fig:dIdVnormal}(a). In Fig.~\ref{Fig:dIdVnormal}(b) we present the theoretically computed $dI/dV$ in the normal state, based on the electronic bandstructure extracted in Ref.~\cite{All13}. The computed $dI/dV$ curve exhibits the same two-peak structure as the experimental data. The two peaks in $dI/dV$ at $E_{1,2}$ arise from the presence of van Hove singularities, which are a direct consequence of the hybridization between the light and heavy bands, as follows from a plot of the equal-energy contours at $E_{1}$ and $E_{2}$ shown in Figs.~\ref{Fig:dIdVnormal}(c) and (d), respectively. Here, the van Hove points are indicated by arrows. We attribute the slight difference in the peak positions between the theoretical and experimental results to different CeCoIn$_5$ samples used in the experiments by Zhou {\it et al.} \cite{Zhou13} and Allan {\it et al.} \cite{All13}. We therefore conclude that the feature in the $dI/dV$ lineshape above $T_c$ reflects the hybridized electronic structure, and in particular the existence of van Hove singularities, in the Kondo screened heavy fermion state, and is unrelated to an electronic pseudogap of the type seen in the cuprate superconductors \cite{Lee06}.

We next turn to the discussion of the differential conductance in the superconducting state. In Fig.~\ref{Fig:SCgap}(b), we reproduce the form of the superconducting gap computed in Ref.~\cite{Dyke14}. We demonstrated that the largest gap is situated on the $\alpha_1$ Fermi surface, with a maximum value of $\Delta^{\alpha_1}_{max}=0.6$ meV, and two smaller gaps with $\Delta^{\alpha_2}_{max}=0.2$ meV and $\Delta^{\beta}_{max}=0.1$ meV exist on the $\alpha_2$ and $\beta$ Fermi surfaces, respectively. As a result, $dI/dV$ in SC state shown in Fig.~\ref{Fig:SCgap}(c) exhibits three sets of coherence peaks corresponding to the maximum SC gaps on the three Fermi surfaces. The peak at ${\bar E}_1$ is a remnant of the van Hove singularity in the normal state at $E_1$ shown in Fig.~\ref{Fig:dIdVnormal}. While the main contribution to $dI/dV$ at small energies arises from the $\beta$-band, with the $dI/dV$ slope thus determined by $\Delta_{\beta}$, the main contribution at higher energies arises from the $\alpha$ band, resulting in much smaller slope in $dI/dV$, and as a result, a kink in $dI/dV$ around $\Delta^{\beta}_{max}$. Note that the non-linear increase of $dI/dV$ at small energies is a direct consequence of the higher-harmonic SC gap in the $\beta$ band $\Delta_{\beta}({\bf k})=\Delta_{\beta}^0 (\cos k_x - \cos k_y)^3$ \cite{Dyke14}.
\begin{figure}
 \includegraphics[width=8.cm]{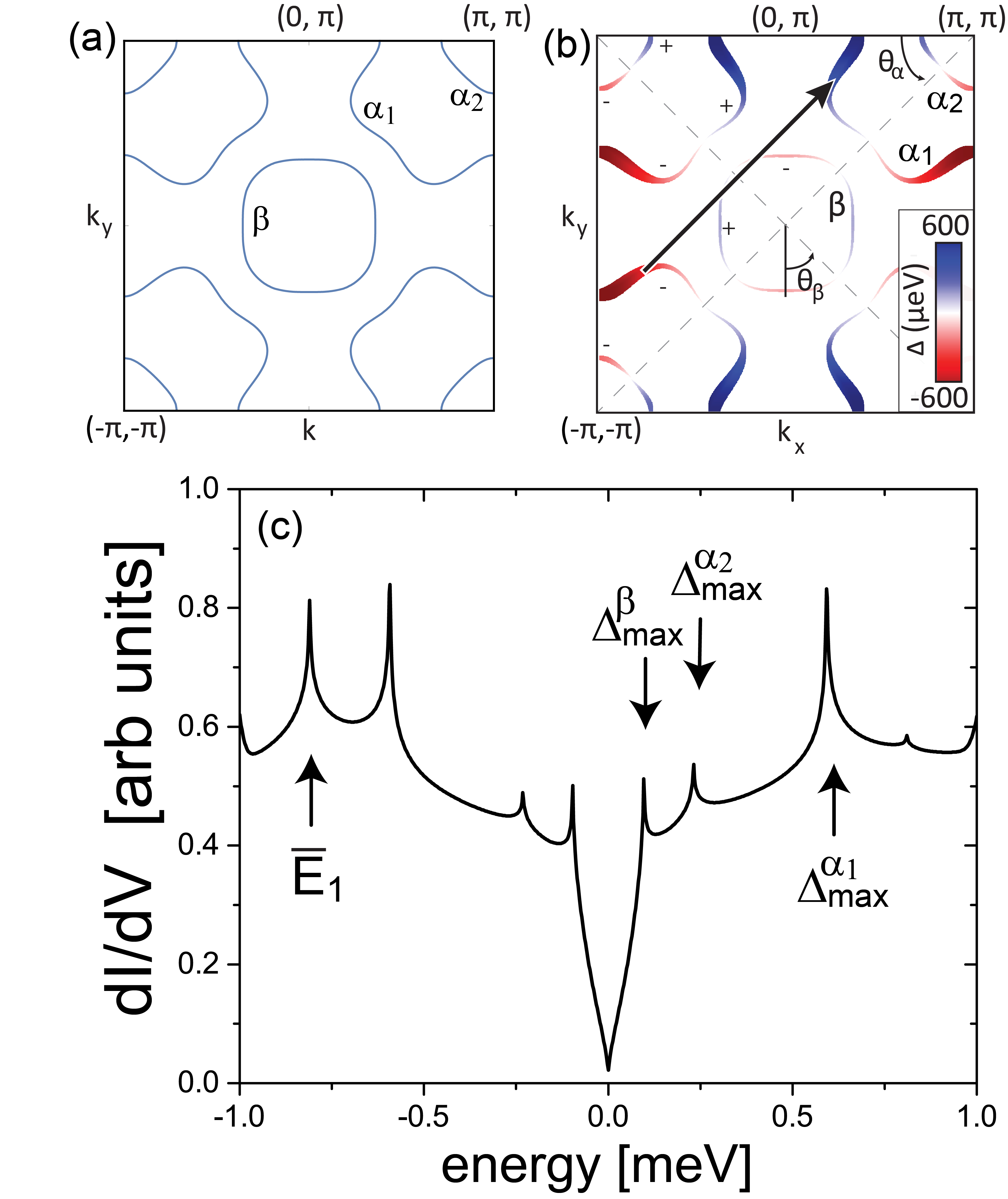}
 \caption{(a) Extracted three Fermi surfaces of the $\alpha$- and $\beta$-bands \cite{All13}. (b) Computed superconducting gap in CeCoIn$_5$ on the $\alpha_1$, $\alpha_2$, and $\beta$ Fermi surfaces \cite{Dyke14}. (c) Theoretically computed $dI/dV$ in the superconducting state. The arrows indicate the position of the coherence peaks associated with the three Fermi surfaces. \label{Fig:SCgap}}
 \end{figure}

\begin{figure}
 \includegraphics[width=8cm]{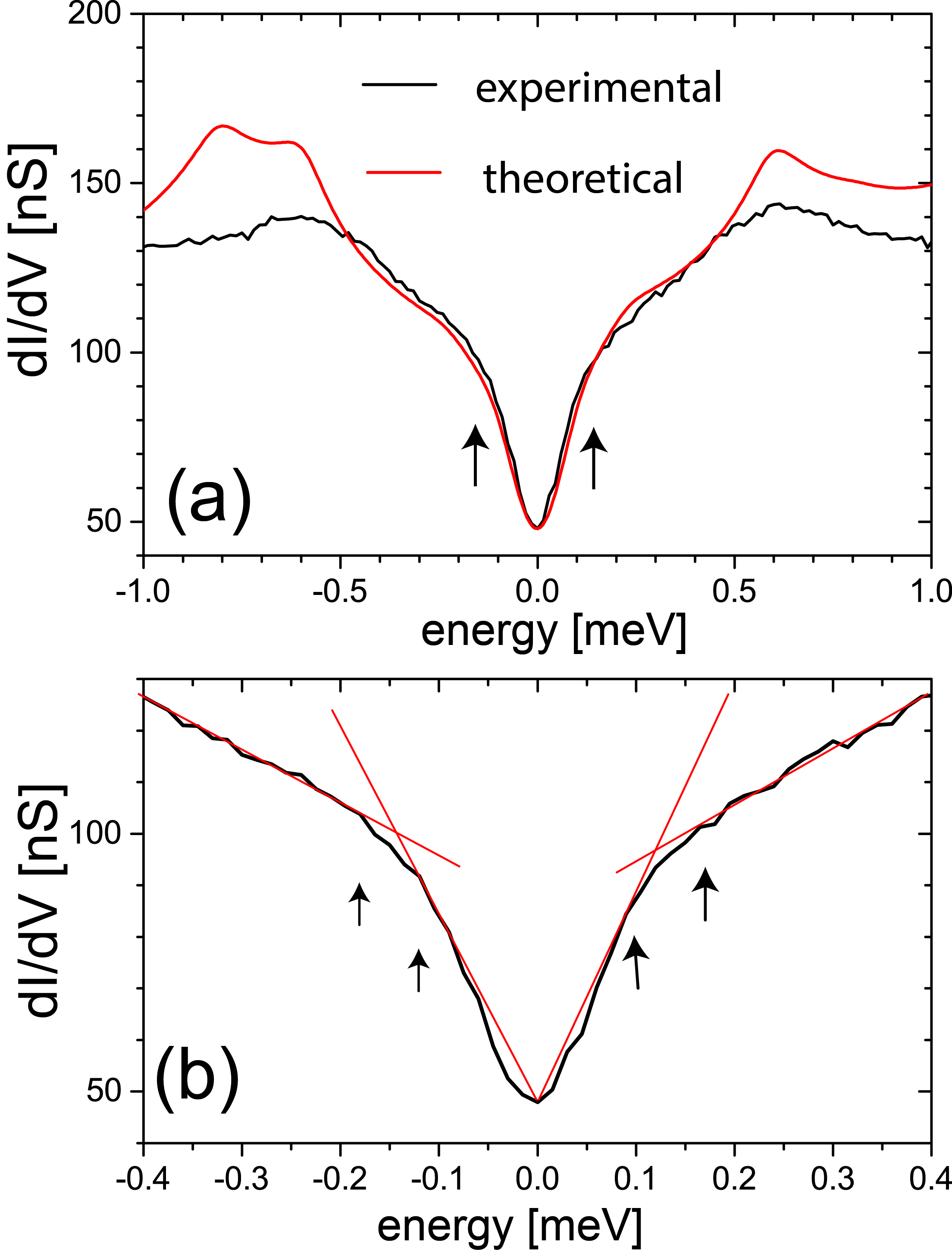}
 \caption{(a) Comparison of the theoretical and experimental \cite{Zhou13} $dI/dV$ curves (courtesy of A. Yazdani) in the superconducting state of CeCoIn$_5$. The theoretical data were broadened by a small quasi-particle damping $\Gamma = 0.06$ meV to account for the experimental resolution, a background was subtracted, and the data were scaled vertically to fit the experimental intensity. (b) Linear fits to the experimental $dI/dV$ curve in the superconducting state.  \label{fig:dIdV_SC_clean}}
 \end{figure}

In Fig.~\ref{fig:dIdV_SC_clean} we present a comparison of the calculated $dI/dV$ with the experimental results of Ref.\cite{Zhou13} for a clean system.  To account for the experimental resolution, we have broadened our results by a small quasi-particle damping $\Gamma = 0.06$ meV. As a result, the coherence peaks at $\Delta_{max}^\beta$ and $\Delta_{max}^{\alpha_2}$ have been smeared out, and have become part of a broader kink in the $dI/dV$ lineshape around $E \approx \pm 0.15$ meV (see arrows). The good agreement between the theoretical and experimental data allows us to conclude that the experimentally observed peaks at $E \approx \pm 0.6$ meV are associated with the larger superconducting gap in the $\alpha_1$-band, while the kink at $E \approx 0.15$ meV reflects the existence of multiple superconducting gaps, and in particular, the unresolved and broadened coherence peaks at $\Delta_{max}^\beta$ and $\Delta_{max}^{\alpha_2}$. Interestingly, using linear fits to the low ($E < 0.1$ meV) and high energy ($0.2 {\rm meV} < E < 0.6$ meV) part of the $dI/dV$ curve, as shown in Fig.~\ref{fig:dIdV_SC_clean}(b), we find that the deviations of the experimental $dI/dV$ curve from linearity (as indicated by arrows) at $E \approx \pm 0.12$ meV and $E \approx \pm 0.18$ meV still agree approximately with the theoretically predicted position of the coherence peaks arising from superconducting gaps in the $\beta$ and $\alpha_2$ bands. Future experiments are clearly required to resolve the coherence peaks associated with these gaps.

We next investigate the changes in the electronic structure arising from the presence of defects, and their corresponding signatures in $dI/dV$ \cite{Liu13}. To account for the scattering of electrons off defects, we employ the $T$-matrix approach, where the unperturbed fermionic Green's function in Matsubara space is given by ${\hat g}({\bf r},{\bf r^\prime}, \tau,\tau^\prime,) = - \langle T_\tau \Psi_{\bf r}(\tau) \Psi^\dagger_{\bf r^\prime}(\tau^\prime) \rangle$ with spinor $\Psi^\dagger_{\bf r} = \left(c^\dagger_{{\bf r}, \uparrow}, c_{{\bf r }, \downarrow}, f^\dagger_{{\bf r}, \uparrow}, f_{{\bf r}, \downarrow} \right)$. We consider a non-magnetic defect located at site ${\bf R}$ which scatters electrons within the $f$- or $c$-bands, such that the full Green's function within the $T$-matrix approach and in Matsubara frequency space is given by
\begin{align}
{\hat G}({\bf r},{\bf r^\prime},i \omega_n) &= {\hat g}({\bf r},{\bf r^\prime},i \omega_n) \nonumber \\
& \hspace{-1.5cm} + {\hat g}({\bf r},{\bf R},i \omega_n) \left[ {\hat 1} - {\hat U} {\hat g}({\bf R},{\bf R},i \omega_n)\right]^{-1} {\hat U} {\hat g}({\bf R},{\bf r^\prime},i \omega_n)
\end{align}
Here,
\begin{equation}
{\hat U} =
\begin{pmatrix}
U_c \sigma_z& 0 \\
0 & U_f \sigma_z
\end{pmatrix}
\end{equation}
with $U_c$ and $U_f$ being the scattering potentials for intraband scattering in
the $c$- and $f$-electron bands, respectively, and $\sigma_z$ is a Pauli matrix.

\begin{figure}
 \includegraphics[width=8cm]{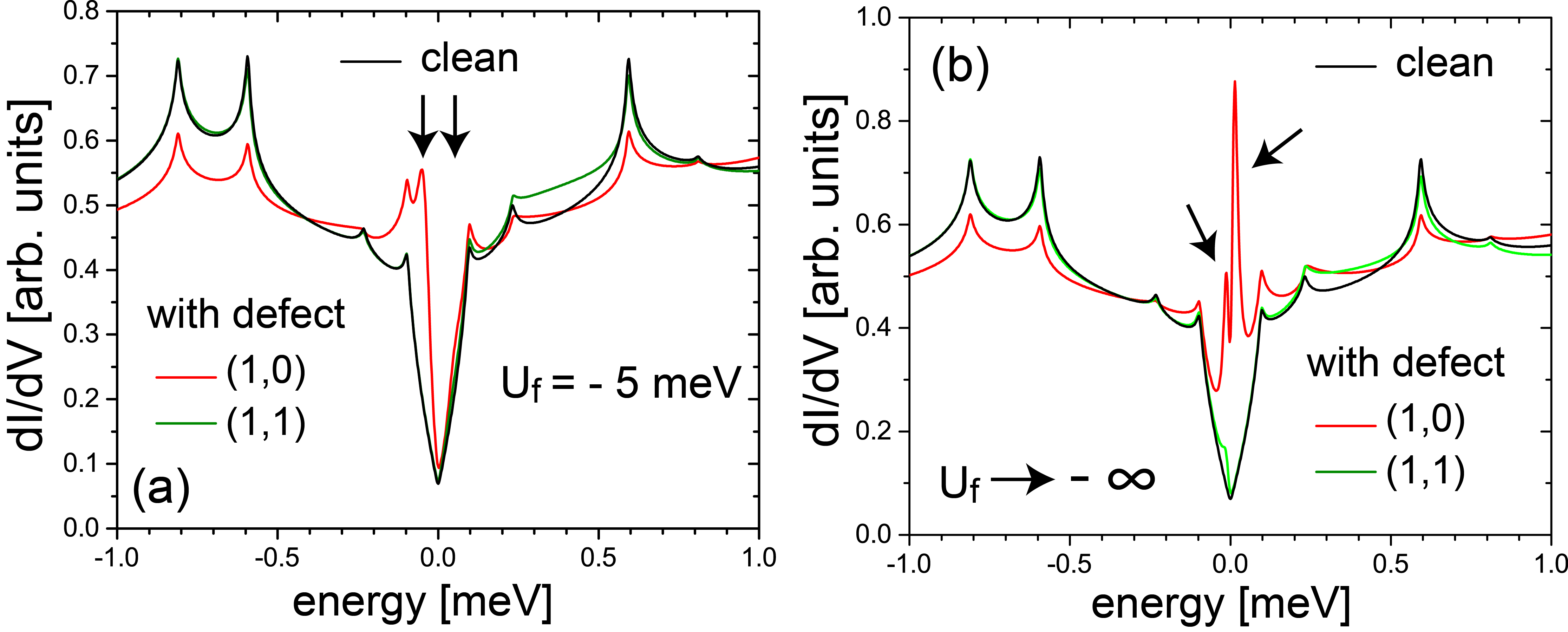}
 \caption{Calculated $dI/dV$ in the vicinity of a potential scatterer in the $f$-band located at ${\bf R}=(0,0)$ for (a) $U_f = -5$ meV, and (b) a Kondo hole ($U_f \rightarrow -\infty$).}
 \label{fig:dIdV_defect}
 \end{figure}

In Fig.~\ref{fig:dIdV_defect}, we present $dI/dV$  in the vicinity of a defect located at ${\bf R}=(0,0)$ for weak scattering $U_f=-5$ meV [Fig.~\ref{fig:dIdV_defect}(a)], and a missing $f$-moment also referred to as a {\it Kondo hole} \cite{Ham11} which we describe via $U_f \rightarrow -\infty$ [Fig.~\ref{fig:dIdV_defect}(b)]. In both cases, the defect induces an impurity state inside the SC gap \cite{Liu13}, as reflected in the presence of additional peaks in $dI/dV$ at positive and negative energies (see arrows). It is interesting to note that even for rather weak scattering of $U_f=-5$ meV, the impurity state is located already well inside the (larger) SC gap at an energy  $E<\Delta_{max}^\beta$. For the case of a Kondo hole, the impurity states are located near zero energy, indicating that a missing $f$-moment acts as a unitary scatterer (the fact that the defect states do not occur exactly at zero energy arises from the particle-hole asymmetry of the system).

\begin{figure}
 \includegraphics[width=8cm]{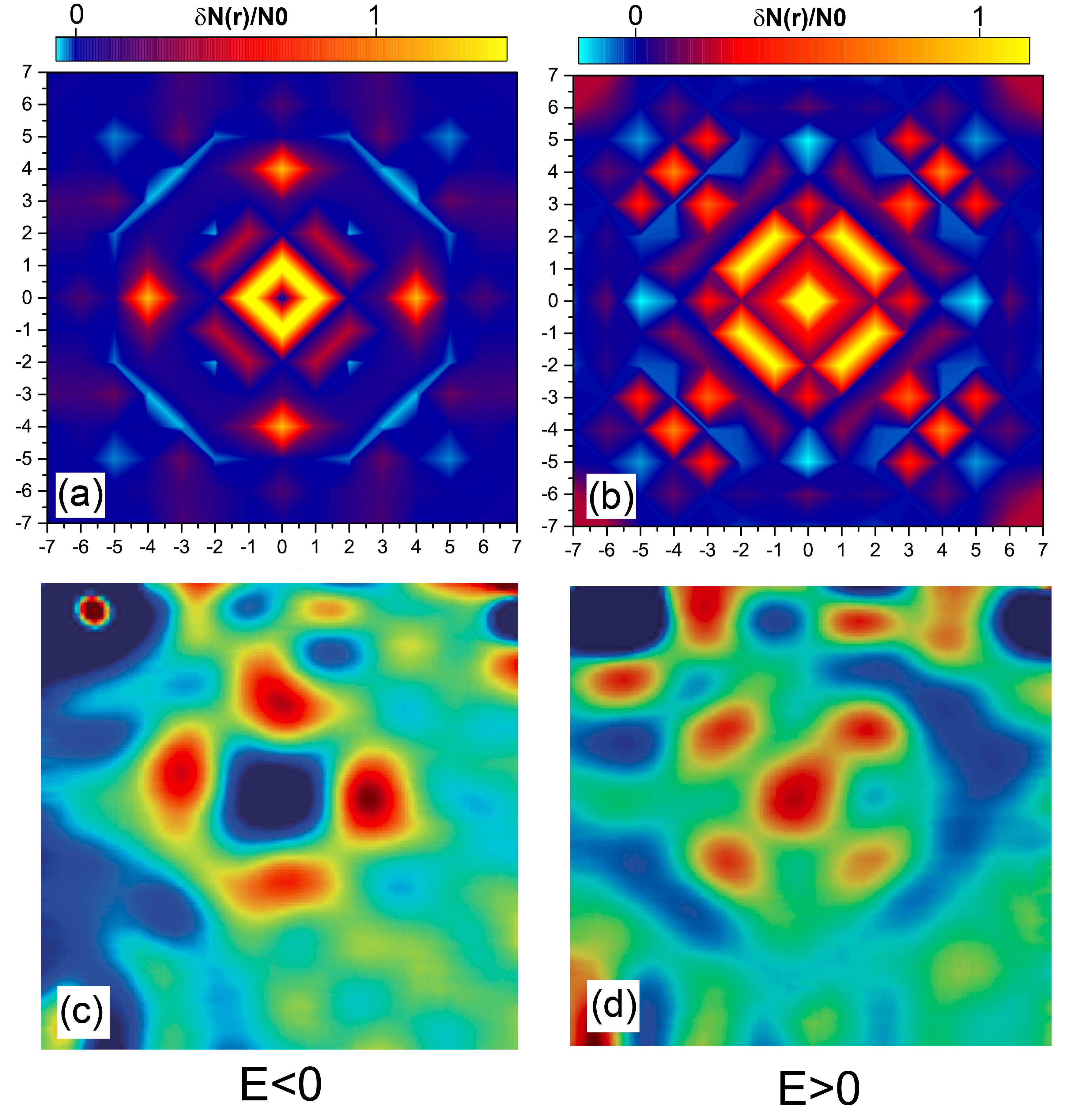}
 \caption{Calculated spatial structure of $dI/dV$ for a scatterer with $U_f=-5$ meV [see Fig.~\ref{fig:dIdV_defect}(a)] located at the center for (a),(b) $E= \mp 0.05$meV. (c) and (d) show the corresponding experimental results from Ref.~\cite{Zhou13}. All four panels show a field of view approximately 14 lattice constants in linear size. }
 \label{fig:dIdV_spatial}
 \end{figure}

The spatial structure of the particle-like and hole-like components of the impurity state \cite{Haas00} yields a characteristic spatial form of $dI/dV$ at the negative and positive energies of the impurity peak positions, as shown in Figs.~\ref{fig:dIdV_spatial}(a) and (b), respectively. The spatial structure of $dI/dV$  between these two energies exhibits a characteristic 45$^\circ$ rotation, as previously observed in the cuprate superconductors which also possess a $d_{x^2-y^2}$-wave symmetry of the superconducting order parameter \cite{Hud01,Haas00}. Our findings are in good agreement with the experimental observations by Zhou {\it et al.} \cite{Zhou13} of a 45$^\circ$ rotation between negative and positive energies, as shown in Fig.~\ref{fig:dIdV_spatial}(c) and (d). The remaining differences between the theoretical and experimental results, such as the larger spectral weight in the vicinity of the defect in Fig.~\ref{fig:dIdV_spatial}(b), could potentially arise from the extended nature of the defect, and a more complicated scattering potential involving inter- and intraband scattering in the $c$- and $f$-bands. The overall good agreement between the theoretical results [Figs.~\ref{fig:dIdV_spatial}(a) and (b)] and the corresponding experimental results [Figs.~\ref{fig:dIdV_spatial}(c) and (d)] provide further evidence for the validity of the electronic bandstructure extracted in Ref.~\cite{All13,Dyke14} and symmetry of the superconducting order parameter computed in Ref.~\cite{Dyke14}.

Finally, we had previously shown \cite{Fig11} that defects in heavy fermion materials lead to spatial oscillations in the hybridization, with the spatial structure of these oscillations reflecting the form of the unhybridized conduction Fermi surface \cite{Ham11}. Work is under way to investigate the effects of these oscillations, and those of the superconducting order parameter, on the spatial and energy structure of the defect-induced impurity states.  While the complex electronic and magnetic structure of CeCoIn$_5$ \cite{Dyke14} renders this investigation computationally very demanding, we found in a simplified model with weaker magnetic interactions, that these spatial oscillations do not lead to any qualitative changes in the $dI/dV$ lineshape or the position of the impurity states.

In conclusion, we have shown that the electronic bandstructure determined from QPI spectroscopy \cite{All13} and the theoretically predicted existence of multiple superconducting gaps \cite{Dyke14} can be used very successfully to explain the experimentally measured $dI/dV$ curves in the normal and
superconducting state of CeCoIn$_5$. We also demonstrated that the pseudo-gap like feature in $dI/dV$ in the normal state is the natural consequence of the existence of van Hove singularities arising from the hybridization of the light and heavy bands. In the superconducting state, the $dI/dV$ lineshape reflects the existence of multiple superconducting gaps. Its form in the vicinity of defects is consistent with a $d_{x^2-y^2}$-symmetry of the superconducting order parameter on all three Fermi surface sheets. The good agreement between our theoretical results and SI-STS experiments \cite{Zhou13} provides further validity for the microscopic pairing mechanism, and the resulting form of the superconducting gaps, proposed in Ref.~\cite{Dyke14}.

\begin{acknowledgments}
We would like to thank M. Allan, F. Massee, A. Yazdani and B. Zhou for helpful discussions, and A. Yazdani and B. Zhou for providing us with the data of Ref.~\cite{Zhou13}.  This work was supported by the U. S. Department of Energy, Office of Science, Basic Energy Sciences, under Award No. DE-FG02-05ER46225 (J.S.V.D. and D.K.M.) and used resources of the National Energy Research Scientific Computing Center, a DOE Office of Science User Facility supported by the Office of Science of the U.S. Department of Energy under Contract No. DE-FG02-05ER46225 (J.S.V.D. and D.K.M.). Experimental contributions to this research were supported by US DOE under contract number DE-AC02-98CH10886 (J.C.D.)

\end{acknowledgments}

\end{document}